\documentclass[10pt]{iopart}
\usepackage{epsf}
\usepackage{iopams}
\usepackage{subfigure}
\begin{document}

\letter{Supporting random wave models: a quantum mechanical approach}

\author{Juan Diego Urbina and Klaus Richter}

\address{Institut f{\"u}r Theoretische Physik,
           Universit{\"a}t Regensburg,
           D-93040 Regensburg, Germany}

\ead{juan-diego.urbina@physik.uni-regensburg.de}

\date{\today}

\begin{abstract}
We show how two-point correlation functions derived within non-isotropic random wave models are in fact quantum results that are obtained in the appropriate limit in terms of the exact Green function of the quantum system. Since no statistical model is required for this derivation,  this shows that taking the wave functions as Gaussian processes is the only assumption of those random wave models. We also show how for clean systems the two-point correlation function defined through an energy average defines a Gaussian theory which substantially reduces the spurious contributions coming from the normalisation problem. 
\end{abstract}
\vspace*{-5mm}

    


%
\vspace{5mm}

Since Berry's seminal paper in 1977 \cite{berr1}, the so-called Random Wave Model (RWM) has become by far the most popular and successful tool to describe the statistical properties of wave functions of classically chaotic systems which in this approach are modelled by a random superposition of plane waves. Its applications range from the realm of optics \cite{berr3}, passing by the general problem of wave mechanics in disordered media \cite{mir1} to important issues in mesoscopic systems \cite{lew1}. Owing to this robustness this approach has been regarded as {\it the} indicator of wave signatures of classical chaotic dynamics \cite{uz1}.

The reasons for this success can be traced back to two fundamental points. First, it can be formally shown that such a random  wave function is a stationary random process \cite{mac} (roughly speaking a function taking random values at each point); second, such random process is Gaussian, namely, it is uniquely characterised by a two-point correlation function which expresses fundamental symmetries, like the isotropy of free space. The fact that the process is Gaussian represents a considerable advantage in an operational sense since it provides us with a set of rules to cope with averages over complicated expressions in the way Wick's theorem and its variants do. At the same time the generality of the random wave two-point correlation makes the theory a remarkably good approximation when the effect of the boundaries can be neglected, like in the case of bulk properties.  

When applied to real quantum systems, however, there remain limitations related to the above mentioned ingredients. Concerning the Gaussian assumption, a formal proof showing that a chaotic wavefunction is indeed a Gaussian process is still lacking. Even more, as noted in \cite{mir2} the Gaussian distribution explicitly contradicts the normalisation condition for the wavefunction. In practical terms this means that, when dealing with statistics beyond the two-point correlation function, the Gaussian distribution produces spurious non physical contributions, and attempts to construct a RWM respecting the normalisation constraint lead to severe mathematical difficulties \cite{har1}. Still this assumption is supported by many arguments based on Random Matrix Theory \cite{lew1}, quantum ergodicity \cite{uz1}, information theory \cite{har1}, and Berry's original semiclassical picture \cite{berr1}. Impressive numerical results also support the conjecture at the level of one-point statistics \cite{rom}, and evidence for higher order statistics is given in \cite{lew1,uz1}. Hence, it is appealing to look for a RWM which minimises the spurious contributions due to the normalisation problem while keeping the wavefunction distribution still Gaussian.

About the isotropic character of the theory, constructing a random superposition of waves satisfying both the Schr\"odinger equation  and boundary conditions turns out to be at least as difficult as solving the full quantum mechanical problem by means of standard techniques. To our knowledge the attempts in the direction of a non-isotropic RWM can only deal with highly idealised boundaries such as an infinite straight wall \cite{berr2}, a linear potential barrier \cite{hell1}, and the edge between two infinite lines enclosing an angle of a rational multiple of $\pi$ \cite{hell2}. Also in \cite{den} a variation of the RWM to include finite size effects is presented. The fact that these approximations already produce non-trivial deviations form the isotropic case is an indication of the importance of the inclusion of arbitrary boundaries. 

Our aim in this communication is twofold: first, we shall show that the mentioned results for the two-point correlation function defining the non-isotropic and finite-size RWM can be derived from quantum mechanical expressions, namely, they are independent of any statistical assumption about the wavefunction. Second, we shall show how for a statistical description of wavefunctions using an energy ensemble average, the spurious contributions coming form the normalisation problem are of order $O(1 / N)$ with $N$ the number of members of the ensemble, making their effect negligible for high energies.

{\it Isotropic and non-isotropic random wave models.} We consider solutions of the Schr\"odinger equation $\left(-\frac{\hbar^{2}}{2m} \nabla^{2}+V(\vec{r})\right)\psi_{n}(\vec{r})=E_{n}\psi_{n}(\vec{r})$  for a closed system where the corresponding classical dynamics is chaotic (in the following we take $2m=1$). The RWM  assumes the statistical description of an ensemble of wavefunctions mimicked  by a random superposition of plane waves with local wavenumber $k(\vec{r})=\sqrt{e-V(\vec{r})} / \hbar$, where $e$ is the mean energy of the states under study. For the sake of comparison we shall focus on the following averages, used for the nodal counting statistics (we follow the notation of Berry \cite{berr2}):
\begin{eqnarray}
B(\vec{r}):=\left\langle \psi(\vec{r})^{2}\right\rangle, && D_{x}(\vec{r}):=\left\langle \left(\frac{\partial \psi(\vec{r})}{\partial x}\right)^{2}\right\rangle, \\
D_{y}(\vec{r}):=\left\langle \left( \frac{\partial \psi(\vec{r})}{\partial y}\right)^{2}\right \rangle &,& K_{y}(\vec{r}):=\left\langle \psi(\vec{r})\frac{\partial \psi(\vec{r})}{\partial y}\right\rangle. \nonumber
\end{eqnarray}
We wish to stress, however, that the RWM, being Gaussian, can deal with far more general averages. For billiard systems the isotropic RWM (denoted by a superscript $i$) is defined by the ensemble
\begin{equation}
\psi^{i}(\vec{r})=\sqrt{\frac{2}{J}}\sum_{j=1}^{J}\cos(kx \cos{\theta_{j}} +ky \sin{\theta_{j}} +\phi_{j}),
\end{equation}
and the average $\langle \ldots \rangle$ is defined by integration over a set of independent random phases $\phi_{j} \in (0,2 \pi]$. We also choose $\theta_{j}=2 \pi j / J$, where the limit $J \to \infty$ is always taken after averaging over the $\phi$'s. Explicit calculation then gives the following results \cite{berr3}:
\begin{equation}
B^{i}(\vec{r})=1,\ D^{i}_{x}(\vec{r})=\frac{k^{2}}{2},\ D^{i}_{y}(\vec{r})=\frac{k^{2}}{2}, \ K^{i}_{y}(\vec{r})=0. 
\end{equation}
These results represent bulk approximations to the real situation, since boundary effects are completely neglected. In order to improve this limitation, the following ensemble of non-isotropic ($ni$) superpositions of random waves was introduced in \cite{berr2} to take into account the effect of a straight infinite boundary at $y=y_{0}$ on which we demand the wavefunction to satisfy Dirichlet (D) or Neumann (N) boundary conditions:
\begin{eqnarray}
\psi^{D}(\vec{r})=\sqrt{\frac{4}{J}}\sum_{j=1}^{J}\sin{k(y-y_{0})}\cos(kx \cos{\theta_{j}}+\phi_{j}),&& \\
\psi^{N}(\vec{r})=\sqrt{\frac{4}{J}}\sum_{j=1}^{J}\cos{k(y-y_{0})}\cos(kx \cos{\theta_{j}}+\phi_{j}).&& 
\end{eqnarray}
With the averaging procedure as in the isotropic case, one obtains for the Dirichlet (upper sign) and Neumann (lower sign) cases \cite{berr2}:
\begin{eqnarray}
B^{ni}(\vec{r})=1\mp J_{0}(2k(y-y_{0})),&& \nonumber \\
D_{x}^{ni}(\vec{r})=\frac{k^{2}}{2}(1\mp J_{0}(2k(y-y_{0}))\mp J_{2}(2k(y-y_{0}))),&& \\ 
D_{y}^{ni}(\vec{r})=\frac{k^{2}}{2}(1 \pm J_{0}(2k(y-y_{0})) \mp J_{2}(2k(y-y_{0}))), && \nonumber \\
K_{y}^{ni}(\vec{r})=\pm kJ_{1}(2k(y-y_{0})).&& \nonumber 
\end{eqnarray}

For more general situations where the confining potential is smooth (S), Bies and Heller \cite{hell1} introduced the following ensemble of random Airy functions ${\rm Ai}(x)$ to satisfy locally the Schr\"odinger equation for a linear ramp potential $V(x,y)=Vy$:
\begin{equation}
\psi^{S}(\vec{r})=\frac{1}{\sqrt{J}}\sum_{j=1}^{J}{\rm Ai}\left[\Psi(y,Q_{j})\right]\exp{\left[i(Q_{j}x+\phi_{j})\right]}.
\end{equation}
Here 
\begin{equation*}
\Psi(y,Q)=\left(\frac{V}{\hbar^{2}}\right)^{\frac{1}{3}}(y-y_{0})+\left(\frac{\hbar^{2}}{V}\right)^{\frac{2}{3}}Q^{2}, 
\end{equation*}
and $y_{0}=e / V$ is the turning point in the direction of the linear ramp, fixed by the mean energy $e$ of the eigenstates under study. The phases $\phi_{j}$ are defined as usual and provide the averaging, while $Q_{j} \in [-\infty,\infty]$. Explicit calculation then gives \cite{hell1,berr4}:
\begin{eqnarray}
B^{S}(\vec{r})=\int_{0}^{\infty}{\rm Ai}^{2}\left[\Psi(y,Q) \right]dQ &&, \nonumber \\
D^{S}_{x}(\vec{r})=\int_{0}^{\infty}Q^{2}{\rm Ai}^{2}\left[\Psi(y,Q)\right]dQ &&, \\ 
D^{S}_{y}(\vec{r})=\int_{0}^{\infty}{\rm Ai}'^{2}\left[\Psi(y,Q)\right]dQ &&, \nonumber \\
K^{S}_{y}(\vec{r})=\int_{0}^{\infty}{\rm Ai}\left[\Psi(y,Q)\right]{\rm Ai}'\left[\Psi(y,Q)\right]dQ && \nonumber 
\end{eqnarray}
where ${\rm Ai}'(x)$ is the derivative of the Airy function.

{\it The quantum description.} We consider a set of normalised solutions ${\psi_{n}(\vec{r})}$ of the Schr\"odinger equation with non-degenerate eigenvalues ${E_{n}}$ lying inside the interval $W=[e-\frac{\delta e}{2},e+\frac{\delta e}{2}]$. Considering W as a range of energies with almost constant mean level spacing $\Delta(e)$, the number of states within the interval is $N=\frac{\delta e}{ \Delta(e)}$ (in general $N_{W}=\int_{e-\frac{\delta e}{2}}^{e+\frac{\delta e}{2}}\rho(E)dE$ with $\rho(E)$ the level density). In the high-energy limit we are interested in, $N \gg 1$ with $\frac{\delta e}{e} \ll 1$ are well defined limits which we shall always assume implicitly. The two point correlation function,
\begin{equation}
F(\vec{r}_{1},\vec{r}_{2}):=\langle \psi(\vec{r}_{1})\psi^{*}(\vec{r}_{2}) \rangle:=\frac{1}{N}\sum_{E_{n} \in W}\psi_n(\vec{r}_{1})\psi^{*}_n(\vec{r}_{2})
\end{equation}
can be used to calculate the averages in Eq. (1) by differentiation:
\begin{eqnarray}
B(\vec{r}):=\left[ F(\vec{r}_{1},\vec{r}_{2})\right]_{\vec{r}_{1}=\vec{r}_{2}=\vec{r}},&& \nonumber \\
D_{x}(\vec{r}):=\left[\frac{\partial^{2}}{\partial x_{1} \partial x_{2}}F(\vec{r}_{1},\vec{r}_{2}) \right]_{\vec{r}_{1}=\vec{r}_{2}=\vec{r}},&& \\ 
D_{y}(\vec{r}):=\left[\frac{\partial^{2}}{\partial y_{1} \partial y_{2}}F(\vec{r}_{1},\vec{r}_{2}) \right]_{\vec{r}_{1}=\vec{r}_{2}=\vec{r}}, && \nonumber \\
K_{y}(\vec{r}):=\left[\frac{1}{2}\left( \frac{\partial}{\partial y_{1}}+\frac{\partial}{\partial y_{2}}\right)F(\vec{r}_{1},\vec{r}_{2})\right]_{\vec{r}_{1}=\vec{r}_{2}=\vec{r}}.&& \nonumber 
\end{eqnarray}
It is convenient to use the Green function of the system,
\begin{equation}
G(\vec{r}_{1},\vec{r}_{2},E+i 0^{+})=\sum_{n=1}^{\infty}\frac{\psi_n(\vec{r}_{1})\psi^{*}_n(\vec{r}_{2})}{E-E_{n}+i 0^{+}},
\end{equation} 
to obtain the expression 
\begin{eqnarray}
F(\vec{r}_{1},\vec{r}_{2})=&& \nonumber \\ \frac{\Delta(e)}{2 \pi i} \frac{1}{\delta e}\int_{e-\frac{\delta e}{2}}^{e+\frac{\delta e}{2}}(G^{*}(\vec{r}_{1},\vec{r}_{2},E+i 0^{+})-G(\vec{r}_{2},\vec{r}_{1},E+i 0^{+}))dE.&&
\end{eqnarray}
Note that this is an exact result and the common approximation $F(\vec{r}_{1},\vec{r}_{2}) \sim (G^{*}(\vec{r}_{1},\vec{r}_{2},e+i 0^{+})-G(\vec{r}_{2},\vec{r}_{1},e+i 0^{+}))$ \cite{sied} {\it is not valid in general and requires further assumptions}. Even more, the additional energy integration in Eq. (12) will turn out to be essential.

Different approximations to the Green function valid under different situations can now be used to study the corresponding wavefunction statistics. 

{\it The bulk contribution and finite size effects.} For billiard systems the bulk ($b$) results are obtained by using the free propagator given in two dimensions by the Hankel function $G^{0}(\vec{r}_{2},\vec{r}_{1},E+i 0^{+})=\frac{i}{4 \hbar}H_{0}^{(1)}(\frac{\sqrt{E}}{\hbar}|\vec{r}_{1}-\vec{r}_{2}|)$ instead of the exact Green function. The corresponding contribution to the two-point correlation is: 
\begin{equation}
F^{b}(\vec{r}_{1},\vec{r}_{2})=\frac{1}{A}\frac{1}{\delta e}\int_{e-\frac{\delta e}{2}}^{e+\frac{\delta e}{2}}J_{0}\left(\frac{\sqrt{E}}{\hbar}|\vec{r}_{1}-\vec{r}_{2}|\right)dE
\end{equation}
where $A$ is the billiard area. Using Eqs. (10) and (13) we easily recover the results in Eq. (3) correctly normalised. Further analysis of  expression (13) shows that it reduces to Berry's result \cite{berr1}
\begin{equation}
 \langle\psi(\vec{r}_{1})\psi(\vec{r}_{2})\rangle^{b}=\frac{1}{A} J_{0}\left(\frac{\sqrt{e}}{\hbar}|\vec{r}_{1}-\vec{r}_{2}|\right)
\end{equation}
when $|\vec{r}_{1}-\vec{r}_{2}| \ll \sqrt{4 A /\pi}$, while it decays much faster for $|\vec{r}_{1}-\vec{r}_{2}| \ge \sqrt{4 A / \pi}$ as long as $\delta e \ge \hbar \sqrt{\pi e /4  A}$. Noticing that $\sqrt{4 A / \pi}$ is just the average system linear size $L$ we see that Eq. (13) actually defines a RWM which incorporates finite size effects when the average is taken on scales larger than the {\it ballistic Thouless energy}, $e_{Th}=\hbar \sqrt{e} /L $. Eq. (13) then provides an analytical expression  for the correlation function defined in \cite{den}.   
 
{\it The case of an infinite straight barrier.} For this situation we construct Green functions with the correct parity under the reflection symmetry with respect to the line $y=y_{0}$ by means of the method of images. The symmetric and antisymmetric  combinations give the corresponding two-point correlation for Dirichlet (upper sign) and Neumann (lower sign) boundary conditions  as:
\begin{eqnarray}
F^{D,N}(\vec{r}_{1},\vec{r}_{2})=\frac{1}{A}\frac{1}{\delta e}\int_{e-\frac{\delta e}{2}}^{e+\frac{\delta e}{2}}\left[J_{0}\left(\frac{\sqrt{E}}{\hbar}\sqrt{(x_{1}-x_{2})^{2}+(y_{1}-y_{2})^{2}}\right) \right. && \nonumber \\
 \left.\pm J_{0}\left(\frac{\sqrt{E}}{\hbar}\sqrt{(x_{1}-x_{2})^{2}+(2 y_{0}-y_{1}-y_{2})^{2}}\right)\right]dE. &&
 \end{eqnarray}
Using this correlation function and Eq. (10) we obtain the averages defined in Eq. (1).  Berry's results Eq. (6) are again obtained in the limit of very short distances to the boundary $|y-y_{0}| \ll \sqrt{\frac{A}{8 \pi}}$.

{\it The infinite, smooth barrier.} For a particle in the potential $V(x,y)=Vy$, the Schr\"odinger equation is separable. The solutions along the $x$ direction are plane waves and in the $y$ direction Airy functions. Hence we have:
\begin{equation}
\psi_{k,e}(x,y)=\left(\frac{8 \pi^{3}}{4 \hbar^{4} V}\right)^{\frac{1}{6}}\exp{(-ikx)} {\rm Ai} \left[\left(\frac{V}{\hbar^{2}}\right)^{\frac{1}{3}}\left(y-\frac{e}{ V}\right)\right]
\end{equation}
and for the Green function
\begin{equation}
G(\vec{r}_{1},\vec{r}_{2},E+i0^{+})=\int_{-\infty}^{\infty}\int_{-\infty}^{\infty}\frac{\psi_{k,e}(\vec{r}_{1})\psi^{*}_{k,e}(\vec{r}_{2})}{E-e-\hbar^{2}k^{2}+i0^{+}}dedk.
\end{equation}
Including this result into Eq. (12) we find for the correlation function:
\begin{eqnarray}
F^{S}(\vec{r}_{1},\vec{r}_{2})=\left(\frac{2 \pi^{3}}{\hbar^{4} V}\right)^{\frac{1}{3}}\frac{1}{N}\int_{e-\frac{\delta e}{2}}^{e+\frac{\delta e}{2}}\int_{0}^{\infty}\cos{(k(x_{1}-x_{2}))} && \\ 
{\rm Ai}\left[\left(\frac{V}{\hbar^{2}}\right)^{\frac{1}{3}}\left(y_{1}-\frac{E-\hbar^{2}k^{2}}{V}\right)\right]{\rm Ai}\left[\left(\frac{V}{\hbar^{2}}\right)^{\frac{1}{3}}\left(y_{2}-\frac{E-\hbar^{2}k^{2}}{V}\right)\right]dkdE &&. \nonumber
\end{eqnarray}
Together with the relations Eq. (10) again, one obtains Eq. (8) in the limit $\Delta(e) \to 0$.

To summarise so far, we obtained the one-point averages, Eq.(1), for a closed system from pure quantum mechanical considerations without appealing to any statistical assumption about the wavefunction. The known RWM results Eqs. (3,6,8) are then derived in the appropriate limits (either short distances or infinite system size). The generalisation to any other average bilinear in the wavefunction is straight forward. 

{\it The normalisation problem.} There is a prominent counter argument against the Gaussian assumption first presented to our knowledge in \cite{mir1} and further explored in \cite{har1}, which deserves special attention. It is the apparent contradiction between the normalisation of the members of the ensemble and the Gaussian distribution of the wavefunction amplitudes. Mathematically this can be stated in the following way. Consider the functional
\begin{equation}
\eta [\psi]=\int|\psi(\vec{r})|^{2}d\vec{r},
\end{equation}
where $\psi(\vec{r})$ is a member of the ensemble we use to describe the statistical properties of the wavefunction. One constraint we must impose is the normalisation of all $\psi(\vec{r})$, expressed by the vanishing of the ensemble variance $Var(\eta)=\langle \left(\eta[\psi]\right)^{2} \rangle -\left(\langle \eta[\psi] \rangle \right)^{2}$. If the wavefunction's distribution is Gaussian, we obtain
\begin{equation}
Var(\eta)=2\int\int |\langle \psi(\vec{r}_{1})\psi^{*}(\vec{r}_{2}) \rangle| ^{2}d\vec{r}_{1}d\vec{r}_{2}.
\end{equation}
This is in clear contradiction to the normalisation condition $Var(\eta)=0$. Here we show that this variance is of order $Var(\eta)=O(1 / N)$. To this end we recall the definition of the two-point correlation, Eq. (9), and note that the $\psi_n(\vec{r})$ are eigenfunctions of the same Hamilton operator, i.e, they form an orthonormal set:
\begin{equation}
\int\psi_{i}(\vec{r})\psi_{j}^{*}(\vec{r})d\vec{r}=\delta_{i,j}.
\end{equation}
Then it is easy to obtain the following composition rule for the two-point correlation:
\begin{equation}
\int \langle \psi(\vec{r}_{1})\psi^{*}(\vec{r}) \rangle \langle \psi(\vec{r})\psi^{*}(\vec{r}_{2}) \rangle d\vec{r}=\frac{1}{N} \langle \psi(\vec{r}_{1})\psi^{*}(\vec{r}_{2}) \rangle.
\end{equation}
Since $\langle \psi(\vec{r}_{1})\psi^{*}(\vec{r}_{2}) \rangle $ converges in the  limit $\frac{e}{\Delta e} \sim {\rm const}$, $N \to \infty$, we see that indeed $Var(\eta)$ converges to zero as $O(1 / N)$.

This behaviour of $Var(\eta)$, relying on the fact that an energy ensemble average is taken, differs from the case of disordered systems where the lack of orthogonality between the different members of the ensemble (since they are eigenfunctions of different Hamiltonians corresponding to different disorder realizations) leads to an expression for $Var(\eta)$ of order $O(1)$.

It is important to note that the extra energy average is essential to satisfy Eq. (23). For example, a simple calculation shows that Berry's result, Eq.\ (14),  {\it does not satisfy the composition rule}, while our result Eq. (13) does it as long as $\delta e \ge e_{Th}$. This is a particular case of a more general statement saying that under certain conditions approximate Green functions will make the correlation function satisfy the composition rule. The proof of this  result requires the use of semiclassical techniques and will be presented elsewhere \cite{yo}.

{\it Concluding remarks.} We have shown that all the two-point correlation functions used to fix the different random wave models, (isotropic, non-isotropic, for a smooth boundary and including finite size effects) can be derived in the appropriate limit of the exact quantum mechanical expressions. {\it To this end we do not use any statistical assumption about the wavefunctions, in fact, this results are independent of the regular, mixed or chaotic character of the classical system} \cite{mar}. Also, we showed that for clean chaotic systems the use of an energy ensemble reduces the spurious contributions coming from the normalisation problem without affecting the Gaussian assumption. This result only requires the consistent use of the quantum mechanical definition of the correlation function. 

A Gaussian field with a correlation given in terms of the exact Green function of the system {\it and the energy average carefully taken into account} represents a generalisation which includes all known RWM as limiting cases and successfully takes into account boundary and normalisation effects for any closed, clean chaotic system. To deal with general shapes (like the ones study in \cite{hell2}) and boundary conditions (like the mixed case presented in \cite{berr4}) , however,  the exact quantum approach presented here cannot be analytically performed. An approach using the semiclassical Green function to derive the asymptotic expressions for the correlations presented here (and the ones in \cite{hell2,berr4}) is the adequate method and will be the subject of a separate communication \cite{yo}.

\ack
We thank Peter Schlagheck, Martin Sieber, and Marko Turek for helpful conversations. JDU is indebted to Sven Gnutzmann, Georg Foltin, and Uzy Smilansky for encouraging and important remarks and acknowledges the kind hospitality at the Weizmann centre of complex systems in Rehovot, Israel, where this work was finished. This work was supported by the Deutsche Forschungsgemeinschaft through the  Graduiertenkolleg "Nonlinearity and Nonequilibrium in Condensed Matter".



\begin{thebibliography}{99}
\bibitem{mar} The reason is that for integrable systems the energy average in Eq. (9) mixes wavefunctions with different symmetries. Even when this can be corrected working with a Green function belonging to the particular subspace under study, the Gaussian assumption is known to fail in this case.
\end{thebibliography}
\end{document}